\documentclass[prd,twocolumn,aps,showpacs,nofootinbib,nobibnotes,superscriptaddress,preprintnumbers]{revtex4}
\usepackage{graphicx,bm,color}
\graphicspath{{./fig/}{./png/}}

\def\blue{\textcolor{black}}
\usepackage{amsmath}

\newcommand{\BoldVec}[1]{\mathchoice%
  {\mbox{\boldmath $\displaystyle     #1$}}%
  {\mbox{\boldmath $\textstyle        #1$}}%
  {\mbox{\boldmath $\scriptstyle      #1$}}%
  {\mbox{\boldmath $\scriptscriptstyle#1$}}%
}
\newcommand{\EQ}{\begin{equation}}
\newcommand{\EN}{\end{equation}}
\newcommand{\EQA}{\begin{eqnarray}}
\newcommand{\ENA}{\end{eqnarray}}

\newcommand{\EEq}[1]{Equation~(\ref{#1})}
\newcommand{\Eq}[1]{Eq.~(\ref{#1})}

\newcommand{\Eqss}[2]{Eqs.~(\ref{#1})--(\ref{#2})}

\newcommand{\Sec}[1]{Sec.~\ref{#1}}
\newcommand{\SSec}[1]{Section~\ref{#1}}

\newcommand{\Fig}[1]{Fig.~\ref{#1}}
\newcommand{\FFig}[1]{Figure~\ref{#1}}
\newcommand{\Tab}[1]{Table~\ref{#1}}
\newcommand{\Figs}[2]{Figs.~\ref{#1} and \ref{#2}}

\newcommand{\bra}[1]{\langle #1\rangle}

{}
{}
{}

\newcommand{\nullvector}{{\bf0}}

\newcommand{\xx}{\BoldVec{x}{}}

\newcommand{\uu}{\BoldVec{u} {}}

\newcommand{\BB}{\BoldVec{B} {}}

\newcommand{\JJ}{\BoldVec{J} {}}

\newcommand{\kk}{\BoldVec{k} {}}

\newcommand{\nab}{\BoldVec{\nabla} {}}

\newcommand{\SSSS}{\bm{\mathsf{S}}}

\newcommand{\emf}{\mbox{\boldmath ${\cal E}$} {}}
\newcommand{\FFF}{\mbox{\boldmath ${\cal F}$} {}}

\def\EEGW{{\cal E}_{\rm GW}}
\def\OmGW{{\Omega}_{\rm GW}}
\def\OmM{{\Omega}_{\rm M}}
\def\OmK{{\Omega}_{\rm K}}
\def\OmT{{\Omega}_{\rm T}}

\def\EErad{{\cal E}_{\rm rad}}

\def\EEcrit{{\cal E}_{\rm crit}}

\def\EM{E_{\rm M}}
\def\ET{E_{\rm T}}

\def\hc{h_{\rm c}}

\def\kf{k_\ast}

\def\tmax{t_{\max}}
\def\kB{k_{\rm B}}

\def\half{{\textstyle{1\over2}}}

\def\onethird{{\textstyle{1\over3}}}

\newcommand{\GeV}{\,{\rm GeV}}

\newcommand{\mHz}{\,{\rm mHz}}

\newcommand{\K}{\,{\rm K}}

\newcommand{\s}{\,{\rm s}}

\newcommand{\cm}{\,{\rm cm}}

\newcommand{\km}{\,{\rm km}}

\newcommand{\Mpc}{\,{\rm Mpc}}

\newcommand{\yan}[4]{, #4, Astron. Nachr. {\bf #2}, #3 (#1).}

\newcommand{\ysci}[4]{, #4, Science {\bf #2}, #3 (#1).}

\newcommand{\ymn}[4]{, #4, Mon.\ Not.\ R.\ Astron.\ Soc.\ {\bf #2}, #3 (#1).}

\newcommand{\yjfm}[4]{, #4, J. Fluid Mech. {\bf #2}, #3 (#1).}

\newcommand{\yprd}[4]{, #4, Phys.\ Rev.\ D {\bf #2}, #3 (#1).}

\newcommand{\yprl}[4]{, #4, Phys.\ Rev.\ Lett.\ {\bf #2}, #3 (#1).}

\newcommand{\yapj}[4]{, #4, Astrophys. J. {\bf #2}, #3 (#1).}

\newcommand{\yapjl}[4]{, #4, Astrophys. J. {\bf #2}, #3 (#1).}

\newcommand{\yanar}[4]{, #4, Astron. Astrophys. Rev. {\bf #2}, #3 (#1).}

\newcommand{\ygafd}[4]{, #4, Geophys. Astrophys. Fluid Dyn. {\bf #2}, #3 (#1).}

\newcommand{\yrpp}[4]{, #4, Rep. Prog. Phys. {\bf #2}, #3 (#1).}

\newcommand{\yjour}[5]{, #5, #2 {\bf #3}, #4 (#1).}

\begin{document}

\title{Numerical simulations of gravitational waves from early-universe turbulence}

\date{\today,~ $ $Revision: 1.334 $ $}
\preprint{NORDITA-2019-024}

\author{Alberto~Roper~Pol}
\email{alberto.roperpol@colorado.edu}
\affiliation{Department of Aerospace Engineering Sciences, University of Colorado, Boulder, Colorado 80303, USA}
\affiliation{Laboratory for Atmospheric and Space Physics, University of Colorado, Boulder, Colorado 80303, USA}
\affiliation{Faculty of Natural Sciences and Medicine, Ilia State University, 3-5 Cholokashvili Street, 0194 Tbilisi, Georgia}

\author{Sayan~Mandal}
\email{sayan.mandal@stonybrook.edu}
\affiliation{McWilliams Center for Cosmology and Department of Physics, Carnegie Mellon University, 5000 Forbes Ave,
Pittsburgh, Pennsylvania 15213, USA}
\affiliation{Faculty of Natural Sciences and Medicine, Ilia State University, 3-5 Cholokashvili Street, 0194 Tbilisi, Georgia}

\author{Axel~Brandenburg}
\email{brandenb@nordita.org}
\affiliation{Nordita, KTH Royal Institute of Technology and Stockholm University, Roslagstullsbacken 23, 10691 Stockholm, Sweden}
\affiliation{Laboratory for Atmospheric and Space Physics, University of Colorado, Boulder, Colorado 80303, USA}
\affiliation{Faculty of Natural Sciences and Medicine, Ilia State University, 3-5 Cholokashvili Street, 0194 Tbilisi, Georgia}
\affiliation{McWilliams Center for Cosmology and Department of Physics, Carnegie Mellon University, 5000 Forbes Ave,
Pittsburgh, Pennsylvania 15213, USA}
\affiliation{Department of Astronomy, AlbaNova University Center, Stockholm University, 10691 Stockholm, Sweden}
\affiliation{JILA, University of Colorado, Boulder, Colorado 80303, USA}

\author{Tina~Kahniashvili}
\email{tinatin@andrew.cmu.edu}
\affiliation{McWilliams Center for Cosmology and Department of Physics, Carnegie Mellon University, 5000 Forbes Ave,
Pittsburgh, Pennsylvania 15213, USA}
\affiliation{Faculty of Natural Sciences and Medicine, Ilia State University, 3-5 Cholokashvili Street, 0194 Tbilisi, Georgia}
\affiliation{Abastumani Astrophysical Observatory, M. Kostava st. 47/57, Tbilisi, GE-0179, Georgia}
\affiliation{Department of Physics, Laurentian University, Ramsey Lake Road, Sudbury, Ontario P3E 2C, Canada}

\author{Arthur~Kosowsky}
\email{kosowsky@pitt.edu}
\affiliation{Department of Physics and Astronomy, University of Pittsburgh, and
Pittsburgh Particle Physics, Astrophysics, and Cosmology Center (PITT PACC), Pittsburgh Pennsylvania 15260}

\begin{abstract}
We perform direct numerical
simulations of magnetohydrodynamic turbulence in the early universe and numerically
compute the resulting stochastic background
of gravitational waves and relic magnetic fields.
These simulations do not make the simplifying assumptions
of earlier analytic work.
If the turbulence is assumed to have an energy-carrying scale that is
about a hundredth of the Hubble radius at the time of generation,
as expected in a first-order phase transition,
the peak of gravitational wave power will be in the mHz frequency range
for a signal produced at the electroweak scale.
The efficiency of gravitational wave (GW) production varies significantly with how
the turbulence is driven.
Detectability of turbulence at the electroweak scale by the planned
Laser Interferometer Space Antenna (LISA) requires anywhere from
\blue{0.5\%} to 10\% of the thermal plasma energy density to be in plasma motions or magnetic fields,
depending on the model of the driving process.
Our results predict a new universal form below the spectral peak
frequency that is shallower than previously thought.
This implies larger values of the GW energy spectra in the low-frequency range.
This extends the range where turbulence is detectable with LISA
to lower frequencies, corresponding to higher energy scales
than the assumed energy-carrying scale.
\end{abstract}
\pacs{98.70.Vc, 98.80.-k}

\maketitle

\section{Introduction}

A period of turbulence in the early universe can produce a stochastic
background of gravitational waves
(GWs).
The turbulence that produces GW radiation might arise from the
dynamics of a first-order phase transition \cite{Witten:1984rs,Hogan:1986qda,Kamionkowski:1993fg},
from the dynamics of primordial magnetic fields \cite{Deryagin},
or from the dynamical coupling
of primordial magnetic fields and the highly conducting primordial plasma
\cite{Brandenburg:1996fc,Christensson:2000sp,Kahniashvili:2010gp,
BKMRPTV17}.
Analytic estimates suggest that turbulence
generated by an electroweak phase transition can produce GWs
within the detectable amplitude and frequency range of the
Laser Interferometer Space Antenna (LISA) if the turbulent
energy density is roughly 1\% of the total energy
density of the Universe at that
time \cite{Kosowsky:2001xp,KGR05,Caprini:2006jb,Gogoberidze:2007an}.

However, the aforementioned analytic estimates
make a number of simplifying assumptions.
Turbulence is assumed to be hydrodynamic
with a typical Kolmogorov
power spectrum and a duration set by
a small fraction of the Hubble time,
omitting the effect of the
expansion of the Universe during the turbulent period.
The inclusion of magnetic fields can
extend the frequency range of the resulting GWs due to the
transfer of power to larger scales \cite{PFL76,Brandenburg:1996fc}.
These turbulence models depend on the temporal correlation function of the
turbulent velocity field, which was assumed in earlier works
and not computed from magnetohydrodynamic (MHD) simulations.
An accurate treatment of these effects is essential for
reliably establishing the spectral shape of the resulting GW background
and its detectability with upcoming detectors \cite{Romano:2016dpx}.
A proper understanding of turbulent sourcing of GWs
is especially relevant for using LISA to constrain
the parameter space of a first-order phase transition \cite{HHRW17}.

If primordial magnetic fields were present during the early universe,
they could dynamically enhance turbulent plasma
motions and serve as an additional
source of GWs \cite{Deryagin,Kahniashvili:2008er,Caprini:2009yp}.
Such magnetic fields can persist until the present epoch.
Lower bounds on the strength of magnetic fields obtained from observations
of TeV blazar spectra \cite{NV10} are suggestive of the existence of these
primordial fields.

Numerical simulations are required to make progress beyond
previous analytic estimates, as pointed out in a recent report
of the LISA cosmology working group \cite{LISA19}.
We present here the results of direct numerical simulations of MHD
turbulence and the resulting stochastic GW spectra.
Given that the turbulent energy densities are below 10\%
of the total energy density of the Universe,
the bulk motions in our simulations are
subrelativistic, but the equation of state is still a relativistic one.
We use the {\sc Pencil Code} \cite{PC}, a sixth-order finite-difference
code using third-order time stepping for the MHD equations and a novel
approach for numerically solving the GW equation, which is
discussed in a separate paper
\cite{Pol:2018pao}.

The present paper is arranged as follows.
\SSec{equations} presents the equations that describe
the production of GWs and the dynamics of
the magnetic and velocity fields during the radiation-dominated
epoch of the early universe.
The initial conditions and the setup of the simulations are
presented in \Sec{simulations}.
The results of the numerical simulations are presented and compared with
previous analytic estimates, and the prospects of detectability with LISA
are discussed in \Sec{results}.

Electromagnetic quantities are expressed in Lorentz-Heaviside units
where $\mu_0=1$. 
Einstein index notation is used, so summation is assumed over repeated
indices.
Latin indices $i$ and $j$ refer to spatial coordinates 1 to 3.

\section{Equations}
\label{equations}

We assume the evolution of the background universe to be described by the spatially
flat, homogeneous, and isotropic
Friedmann-Lema\^itre-Robertson-Walker metric
$g_{ij} =  a^2 \delta_{ij}$,
with $a$ being the scale factor.
The expansion of the Universe described by the temporal evolution of
$a$ leads to a dilution of radiation energy
density and magnetic fields, and to the damping of the GW amplitude.
It is convenient to scale out the effects of expansion
by using conformal time $t$, defined as $dt = dt_{\rm phys}/a$, and
comoving coordinates $\xx = \xx_{\rm phys}/a$ and MHD fields.
The physical coordinates and time are expressed as
$\xx_{\rm phys}$ and $t_{\rm phys}$.
The ultrarelativistic equation of state $p = \rho c^2/3$ is valid during
radiation domination.
This leads to a linear evolution of the
scale factor with conformal time $t$ as the solution
to the Friedmann equations \cite{Friedmann1922}.

\subsection{Gravitational wave equation}
\label{GW_eq}

We consider small tensor-mode perturbations $a^2 h_{ij}^{\rm phys}$ over the
background metric $g_{ij}$.
The GW equation is then \cite{Gri74, Mukhanov}
\begin{align}
\biggl( \partial^2_{t_{\rm phys}} + 3 H (t) \partial_{t_{\rm phys}} -
c^2 \nab^2_{\text{phys}} \biggr) h_{ij}^{\rm phys} (\xx, t)
\nonumber \\ = \frac{16 \pi G}{c^2} T^{\text{TT}}_{ij, \text{phys}} (\xx, t),
\label{GW1}
\end{align}
where $c$ is the speed of light, $G$ is Newton's gravitational constant, and
$H = a^{-2}\partial_t a$ is the Hubble rate.
The transverse and traceless stress tensor $T_{ij, {\rm phys}}^{\rm TT}$ 
sources the gauge-free metric perturbations \cite{MTW73}.

The introduction of comoving coordinates, conformal time,
comoving stress tensor
$T_{ij}^{\rm TT} = a^4 T_{ij, {\rm phys}}^{\rm TT}$,
and scaled strains $h_{ij} = a h_{ij}^{\rm phys}$ simplifies
\Eq{GW1} to \cite{Pol:2018pao}
\begin{equation}
\left( \partial_t^2 - c^2 \nab^2 \right)
h_{ij}(\xx, t) = \frac{16 \pi G}{a c^2} T_{ij}^{\text{TT}} (\xx, t),
\label{GW3}
\end{equation}
where the omitted damping term $h_{ij}\, a^{-1}\partial_t^2 a$ is zero during
radiation-domination.

The stress tensor $\bar{T}_{ij}^{\rm TT}$ is normalized by the 
energy density at the initial time of turbulence
generation $t_*$, which,
during the radiation-dominated era, is
\begin{equation}
\EEcrit^\ast = \frac{3 H_*^2 c^2}{8 \pi G} \approx
\EErad^* = \frac{\pi^2 g_* (k_{\rm B} T_*)^4}{30 \, (\hbar c)^3},
\label{rho_rad}
\end{equation}
where $T_*$, $g_*$, and $H_\ast$ are the temperature, the number
of relativistic degrees of freedom, and the Hubble rate, respectively,
$k_{\rm B}$ is the Boltzmann constant,
and $\hbar$ is the reduced Planck constant.

We also normalize conformal time by $t_*$ and define $\bar{t} = t/t_*$.
Because of the linear evolution of the scale factor $t_\ast = H_\ast^{-1}$,
where $a_\ast = 1$ has been used.
The scale factor is just $a = \bar{t}$.
Note that in this convention, $a_0 \neq 1$ at the present time.
Similarly, the space coordinates are normalized as
$\bar\xx = \xx H_\ast/c$.
\EEq{GW3} reduces then to the normalized GW equation,
\begin{equation}
\left( \partial^2_{\bar{t}} - \bar\nab^2\right)
h_{ij} (\bar{\xx}, \bar{t})\, =\, {6 \over \bar{t}} \bar{T}_{ij}^{\rm TT}
(\bar{\xx}, \bar{t}).
\label{GW4}
\end{equation}
This is the wave equation that we solve within the {\sc Pencil Code}.
From now on, we omit overbars and always refer to normalized variables,
unless stated otherwise.

The use of normalized variables has the advantage
that our simulations can readily be applied to different
energy scales where our equations are applicable, i.e., after the
electroweak phase transition and within
the radiation-dominated epoch.
Through a simple set of steps, the final diagnostic frequency diagram
can then easily be assembled from the normalized results.

\subsection{Gravitational wave characteristics}

The characteristic amplitude of GWs is defined as \cite{Mag00}
\begin{equation}
h_c^2(t) = \frac{1}{2} \bigl\langle \bigl(h_{ij}^{\rm phys}(\xx, t)\bigr)^2
\bigr\rangle =
\frac{1}{2 t^2} \bigl\langle h^2_{ij} (\xx, t)\bigr\rangle,
\label{hc}
\end{equation}
where angle brackets denote averaging over the physical volume,
and the second equality is true during the radiation-dominated epoch
with the normalization described above.
The spectrum of the characteristic amplitude is defined
following Ref.~\cite{Mag00}, such that
$\int \hc^2 (k, t) \, d \ln k = \hc^2 (t)$;
see details in Ref.~\cite{Pol:2018pao}.
The integration to compute $\hc$ is performed over wave numbers $k$
from 0 to $\infty$.
Since the spectral function $\hc(k)$ is defined to be integrated
in $\ln k$, the limits of integration become
$-\infty$ to $\infty$.
Note that $k$ refers to the normalized wave number $\bar k$,
consistently given by $\bar k = c k/H_\ast = a c k_{\rm phys}/H_\ast$,
where again we omit the overbar from now on.

In the absence of turbulent sources, and neglecting details of the GW transfer
function due to evolving relativistic degrees of freedom and transitions
between radiation, matter, and dark energy dominations (see Ref.~\cite{SS18}),
the characteristic amplitude $\hc (k, t)$
dilutes due to the expansion of the Universe as $a^{-1}$.
Hence, the relic observable
amplitude at the present time $\hc (k)$ is the amplitude at the end of the
simulation $t_{\rm end}$
diluted by a factor $t_{\rm end}/a_0$.
Note that $t_{\rm end}$ is assumed to be within the radiation-dominated epoch,
such that the computed numerical results and the described normalization are valid.
The value of the scale factor $a_0$ is obtained assuming adiabatic expansion of the
Universe, i.e., such that $g_{\rm S}\, T^3 a^3$ stays constant,
where $g_{\rm S}$ is the number
of adiabatic degrees of freedom.

The physical energy density carried by the GWs $\EEGW (t)$
is defined as \cite{Mag00}
\begin{equation}
\EEGW (t) = \frac{c^2}{32 \pi G} \bigl\langle \bigl(\partial_{t_{\rm phys}}
h_{ij}^{\rm phys} (\xx, t)\bigr)^2 \bigr\rangle,
\label{EEGW}
\end{equation}
which we normalize by the radiation energy density,
$\OmGW (t) = \EEGW (t)/\EErad^\ast$.
In terms of conformal time $t$ and scaled strains $h_{ij} (\xx, t)$
during the radiation-dominated epoch, our normalization leads to
\begin{equation}
\OmGW (t) = \frac{1}{12 t^4} \bigl\langle \bigl(\partial_t h_{ij} (\xx, t) -
h_{ij} (\xx, t)/t\bigr)^2 \bigr\rangle.
\label{Om}
\end{equation}
The GW energy density spectrum $\OmGW (k, t)$ is defined as in
Ref.~\cite{Mag00}, such that $\int
\OmGW (k, t)\,d \ln k = \OmGW (t)$; see details in
Ref.~\cite{Pol:2018pao}.
This is the standard normalization
that we use within the {\sc Pencil Code}.
However, when we are
interested in the observable relic signal,
it is useful
to normalize by the critical energy density at the present time,
$\EEcrit^0 = (3 H_0^2 c^2)/(8 \pi G)$,
where $H_0 = 100 \, h_0 \km\, \s^{-1} \Mpc^{-1} \approx 3.241 \times 10^{-18} \, h_0 \s^{-1}$
is the Hubble rate at the present time.
We use $h_0^2 \OmGW$ to quote our results independently of the uncertainties
in the value of $h_0$ \cite{Mag00}.
The GW energy density dilutes due to the expansion of the Universe as $a^{-4}$.
Hence, the relic observable at the present time
$\OmGW (k)$ is the energy density in \Eq{Om} at the end of the simulation
$t_{\rm end}$, reduced by a factor $(H_\ast/H_0)^2 (t_{\rm end}/a_0)^4 \sim g_\ast^{-1/3}$.
During the radiation-dominated epoch, $a^4 \OmGW$ is constant.
However, since $a$ does not evolve as $t$ all the way to the present time,
this factor is not unity.
Finally, we express the GW amplitude $\hc(f)$
and the energy spectra $\OmGW (f)$, which are observables
at the present time---as a function of the physical frequency, shifted to the present
time.
The frequency is related to $\bar k$ through
\begin{equation}
f = \frac{c k_{\rm phys}}{2 \pi} = \frac{H_\ast a_0^{-1}}{2 \pi} \bar k.
\label{freq}
\end{equation}

\subsection{MHD equations}
\label{MHD_eqs}

The GW equation is sourced by the stress tensor $T_{ij}$.
In particular, we consider GWs sourced by MHD turbulence.
Starting with initial conditions for the plasma velocity and
magnetic fields at the starting time of the turbulence period, we
numerically solve for the dynamics of early-universe MHD turbulence
using the {\sc Pencil Code}.
At each time step, we compute
the spatial Fourier components of the stress tensor of a
relativistic perfect fluid,
\begin{equation}
T_{ij} =\frac{4}{3}\frac{\rho u_i u_j}{1 - \uu^2} - B_i B_j + (\rho/3 + \BB^2/2)
\delta_{ij},
\label{Tij}
\end{equation}
where $\uu$ is the plasma velocity and $\BB$ is the magnetic field.
The MHD fields $\rho = a^4 \rho_{\rm phys}$ and
$\BB = a^2 \BB_{\rm phys}$ are expressed as comoving variables.
We emphasize that in MHD, the Faraday displacement current is omitted.
This implies that electric fields do not contribute to the stress tensor
\cite{Brandenburg:1996fc, Subramanian15}.

The MHD equations for an ultrarelativistic gas in a flat expanding
universe in the radiation-dominated era after the electroweak phase
transition are given by \cite{Brandenburg:1996fc,BKMRPTV17,DN13}
\begin{eqnarray}
{\partial\ln\rho\over\partial t}
&=&-\frac{4}{3}\left(\nab\cdot\uu+\uu\cdot\nab\ln\rho\right) \nonumber \\
&&+{1\over\rho}\left[\uu\cdot(\JJ\times\BB)+\eta \JJ^2\right],
\label{dlnrhodt}\\
{\partial\uu\over\partial t}&=&-\uu\cdot\nab\uu
+{\uu\over3}\left(\nab\cdot\uu+\uu\cdot\nab\ln\rho\right) \nonumber \\
&&-{\uu\over\rho}\left[\uu\cdot(\JJ\times\BB)+\eta \JJ^2\right]-{1\over4}\nab\ln\rho \nonumber \\
&&+{3\over4\rho}\JJ\times\BB+{2\over\rho}\nab\cdot\left(\rho\nu\SSSS\right)+\FFF,
\label{dudt} \\
{\partial\BB\over\partial t}&=&\nab\times(\uu\times\BB-\eta\JJ+\emf),
\label{dAdt}
\end{eqnarray}
where ${\sf S}_{ij}=\half(u_{i,j}+u_{j,i})-\onethird\delta_{ij}\nab\cdot\uu$
are the components of the rate-of-strain tensor with commas denoting partial
derivatives, $\nu$ is the kinematic viscosity,
and $\eta$ is the magnetic diffusivity.
Energy can be injected into velocity and magnetic fields
through ponderomotive and electromagnetic forces
$\FFF$ and $\emf$, respectively.

All variables are normalized with the appropriate powers
of the radiation energy density and the Hubble rate, both
at the time of generation:
$\bar{\rho} = \rho c^2/\EErad^\ast$, $\bar{\uu} = \uu/c$,
$\bar{\JJ} = (c/H_*) \, \JJ/\sqrt{\EErad^\ast}$, $\bar{\BB} = \BB/\sqrt{\EErad^\ast}$,
$\bar{\eta} = H_* \eta /c^2$, $\bar{\nu} =  H_* \nu /c^2$,
$\bar{\FFF} = \FFF/(H_\ast c)$, and $\bar{\emf} = \emf/\sqrt{c^2 \EErad^\ast}$, where
the overbars have been dropped on \Eqss{Tij}{dAdt}.
In addition, similar to $\rho$ and $\BB$,
the current density $\JJ$ is comoving, i.e., scaled with $a^3$.
The physical value of the magnetic diffusivity $\eta$ at the electroweak
phase transition is given in Eq.~(9) of Ref.~\cite{BSRKBFRK17}: $\eta \approx
4 \times 10^{-9} \, (\kB T_\ast/100 \GeV)^{-1} \cm^2/\s$.
This corresponds to $9.2 \times 10^{-20}$ in our normalized units.

The energy densities of the magnetic and velocity fields are computed as
$\OmM (t) = \bra{\BB^2}/2$ and $\OmK (t) = \bra{\rho \uu^2}/2$.
We define the magnetic and kinetic energy spectrum such that
$\int \Omega_{\rm M, K} (k, t) \, d \ln k =
\Omega_{\rm M, K} (t)$.
Here, $\Omega_{\rm M, K} (k, t)$ are
the spectra in terms of logarithmic wave number intervals.
They are defined analogously to $\OmGW (k, t)$;
see Refs.~\cite{Mag00,DC03,Pol:2018pao}.

\section{Numerical simulations}
\label{simulations}

To compute the resulting GW production,
we evolve the strains in Fourier space using \Eq{GW4},
assuming a constant source term during the length of one time
step of the MHD evolution.
This assumption is accurate for time steps small enough to guarantee
numerical stability of the MHD equations,
and it allows much longer time steps than what is required
by a direct numerical simulation;
see Sec.~2.6 of Ref.~\cite{Pol:2018pao} for a discussion of this new
method, which is described there as approach~II.

It turns out that the GW energy production ceases some time after the
kinetic and/or magnetic energies have started to decay.
The GW spectrum is then statistically steady.
We continue our simulations to gather sufficient statistics to compute
accurate averages for the GW spectra in
comoving variables (more specifically, until the GW spectra start to fluctuate
around a steady state and we have computed at least one period).
In all our simulations, this occurs well within the
radiation-dominated era.
The last time of the numerical simulations
is what we have previously called $t_{\rm end}$.

To study the sensitivity to initial conditions, we have performed several
sets of simulations with different physical models for driving plasma motions.
The motivation for the different types of primordial magnetic
fields obtained below is given in Ref.~\cite{BKMRPTV17}, where
their subsequent evolution and observational constraints are discussed.
The physical magnetic diffusivity $\eta$ of the early universe is
much smaller than what we can accurately simulate.
For similar reasons, thermal and radiative diffusion were not
included in the equations above.
We fix the viscosity
$\nu=\eta$ and choose it to be as small as possible,
but still large enough such that the inertial range of the
computed spectra is appropriately resolved \cite{BKMRPTV17}.
If the much smaller physical values were used
instead, much larger numerical resolution would be required and,
the inertial range of the turbulence would extend to higher frequencies.
Those higher frequencies are of little observational interest since the GW amplitude at those
frequencies would be very low, as seen from our spectra shown below.

\begin{table}[b!]\caption{
Summary of runs.}
{\scriptsize
\vspace{12pt}\centerline{\begin{tabular}{lccccccrc}
Run & ${\cal E}_0, {\cal F}_0$ & $\Omega_i^{\rm max}$ &
$\Omega_{\rm GW}^{\rm sat}$ & $i$ & hel & $t_{\max}$ & \ $N\;$ & $\eta$ \\ 
\hline
ini1 &  ---   & $1.16\times 10^{-1}$ & \blue{$3.38\times 10^{-8}$} & M & y & 1.00 &
	100 & $5\times 10^{-6}$ \\ 
ini2 &  ---   & $7.62 \times 10^{-3}$ & \blue{$1.02\times 10^{-10}$} & M & y & 1.00 & 
	100 & $5 \times 10^{-8}$ \\ 
ini3 &  ---   & $7.62\times 10^{-3}$ & \blue{$9.97\times 10^{-9}$} & M & y & 1.00 &
	10 & $5\times10^{-7}$ \\ 
hel1 & $1.4\times 10^{-3}$ & $2.17\times 10^{-2}$ & $4.43\times 10^{-9}$ & M & y & 1.10 &
	100 & $5\times 10^{-7}$ \\ 
hel2 & $8.0\times 10^{-4}$ & $7.18\times 10^{-3}$ & $4.67\times 10^{-10}$ & M & y & 1.10 &
	100 & $5\times 10^{-7}$ \\ 
hel3 & $2.0\times 10^{-3}$ & $4.62\times 10^{-3}$ & $2.09\times 10^{-10}$ & M & y & 1.01 &
	100 & $5\times 10^{-7}$ \\ 
hel4 & $1.0\times 10^{-4}$ & $5.49\times 10^{-3}$ & $1.10\times 10^{-11}$ & M & y & 1.01 &
	1000 & $2\times 10^{-6}$ \\ 
noh1 & $1.4\times 10^{-3}$ & $1.44\times 10^{-2}$ & $3.10\times 10^{-9}$ & M & n & 1.10 &
	100 & $5\times 10^{-7}$ \\ 
noh2 & $8.0\times 10^{-4}$ & $4.86\times 10^{-3}$ & $3.46\times 10^{-10}$ & M & n & 1.10 &
	100 & $2\times 10^{-6}$ \\ 
ac1  &  3.0   & $1.33\times 10^{-2}$ & $5.66\times 10^{-8}$ & K & n & 1.10 &  100 &
	$2\times 10^{-5}$ \\ 
ac2  &  3.0   & $1.00\times 10^{-2}$ & $3.52\times 10^{-8}$ & K & n & 1.10 &
	100 & $5\times 10^{-5}$ \\ 
ac3  &  1.0   & $2.87\times 10^{-3}$ & $2.75\times 10^{-9}$ & K & n & 1.10 &
	100 & $5\times 10^{-6}$ \\ 
\label{Tsummary}\end{tabular}}}\end{table}

Our full set of runs is summarized in \Tab{Tsummary}.
For all of the calculations,
we assume $\uu(\xx)=\bm{0}$ initially.
In set~I (runs~ini1--3), $\BB (\xx)$ is initialized as a fully helical
(indicated by ``y'' under ``hel'') Gaussian random
field with magnetic energy spectrum
$\Omega_{\rm M} (k)\propto k^5$
for $k<\kf$, corresponding to a solenoidal causally generated field,
and $\Omega_{\rm M}(k) \propto k^{-2/3}$ (Kolmogorov spectrum) for $k>\kf$
where $k_\ast$ is the
wave number at which the magnetic energy is injected.
In set~II (runs~hel1--4 and noh1--2), $\BB(\xx)=\bm{0}$ initially,
but it is then numerically driven by applying
an electromotive force $\emf$
during $1\leq t\leq t_{\max}$ in the induction \Eq{dAdt}
consisting of random and nearly monochromatic waves around wave number $\kf$.

The driving
force field is taken as either fully helical (runs~hel1--4) or nonhelical
(noh1--2), using a forcing term quantified by ${\cal E}_0$
described in Refs.~\cite{Bran01,BDS02}; see Table~\ref{Tsummary}
for values of ${\cal E}_0$ and $\tmax$.
The initial number of eddies per horizon
length at the driving scale is $N\equiv \kf/2\pi$,
usually taken to be between 1 and 100 for the
first-order electroweak phase transition \cite{TWW92}.

We compute the decaying MHD turbulent motions, with no forcing term,
for times $t > t_{\max}$, where $t_{\max}$ has been defined for set~II of runs,
and it can be assumed to be $t_{\max} = 1.0$ for
the runs corresponding to set~I.
We arrange the simulations such that the maximum
total magnetic energy density $\OmM^{\rm max}$
integrated over all wave numbers
is a specified fraction of the radiation energy density.
We take values in the range $10^{-3}$ to $10^{-1}$.
The lower limit is required for the turbulence to be the dominant source of
GWs during a first-order phase transition according to analytic
estimates \cite{Nicolis:2003tg}, and the higher limit is imposed on magnetic fields
due to their effect on big bang nucleosynthesis
\cite{Kahniashvili:2009qi,Yamazaki:2012jd}.
Recently, values up to 10\% have been obtained for magnetogenesis
lattice simulations \cite{ZVF19}.

We also consider set~III (runs~ac1--3) with
initial $\BB (\xx) = \uu(\xx) = \nullvector$,
using irrotational or ``acoustic'' hydrodynamic turbulence.
In this case, the forcing $\FFF$ appears
as an additional term in the momentum \Eq{dudt},
which acts during $1 \leq t \leq \tmax$.
The forcing term, with amplitude ${\cal F}_0$,
is computed as the gradient of a combination of Gaussian
random potentials $\phi \propto
\exp[-(\xx-\xx_i)^2/R^2]$ centered at random positions $\xx_i$
of the domain.
This results in a number of eddies $N=(\pi R)^{-1}$ \cite{MB06}.

We choose solenoidal and irrotational forcing fields in
sets~II and III, respectively, for comparison purposes.
In all of our runs, the size of the cubic domain $L$ is taken to be
$L = 2\pi/N$, such that the lower wave number in the computed
spectra corresponds to $N$.
We evolve the dynamical equations on a mesh of $1152^3$ grid points.

\section{Results}
\label{results}

\subsection{The turbulent GW energy spectrum}
\label{NormalizedResults}

\begin{figure}[t!]\begin{center}
\includegraphics[width=\columnwidth]{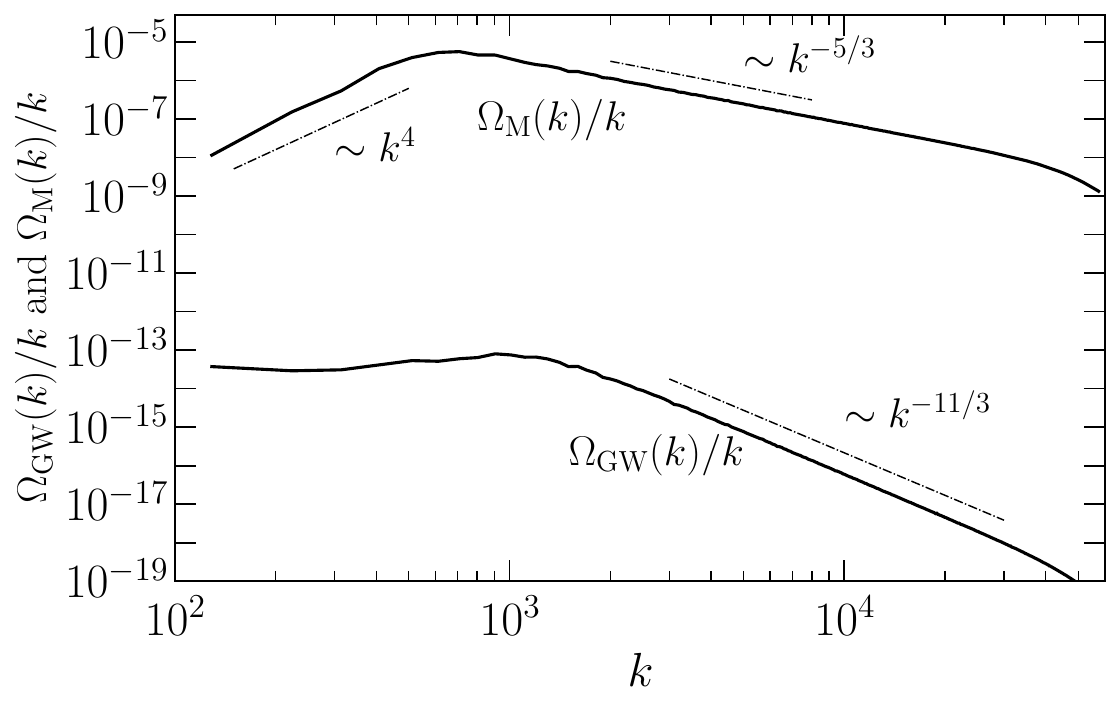}
\end{center}\caption[]{
Magnetic and GW energy spectra for
run~ini2 averaged over late times ($t > 1.1$), after the GW
spectrum has started to fluctuate around a steady
state.
}\label{pspecm_M1152e_exp6k4}\end{figure}

We begin by discussing the spectrum for a run of set~I.
\FFig{pspecm_M1152e_exp6k4} shows the resulting magnetic field and
GW energy spectra for such a case with a Kolmogorov-type spectrum.
In this case, the magnetic field has a Batchelor ($k^4$) spectrum
in the subinertial range and a Kolmogorov ($k^{-5/3}$) spectrum
in the inertial range.
For the resulting stress, this corresponds to a white noise ($k^2$)
spectrum in the subinertial range and to the same Kolmogorov power law in the
inertial range; see the Appendix and Ref.~\cite{BB19}.
The GW energy density shows a spectrum proportional to $k^{-2}$
with respect to the spectrum of the stress.
This can be understood by comparing the order of $k$ of the different terms
in \Eq{GW1}.
The third and fourth terms of the equation in Fourier space
are $c^2 k_{\rm phys}^2 \tilde{h}^{\rm phys}_{ij} (\kk, t)$, and
$(16 \pi G/c^2) \tilde{T}_{ij, {\rm phys}}^{\rm TT} (\kk, t)/t$.
Therefore, if one assumes these terms to be of the same order, then
$k^4 \tilde{h}_{ij}^{\rm phys} (\kk, t)
\tilde{h}_{ij}^{\rm phys} (\kk, t)
\sim \tilde{T}_{ij, {\rm phys}}^{\rm TT} (\kk, t)
\tilde{T}_{ij, {\rm phys}}^{\rm TT} (\kk, t)/t^2$.

We shell integrate both sides to obtain a term proportional
to the spectrum,
$k^4 \hc^2(k, t) \sim \OmT (k, t)/t^2$.
We define the stress spectrum $\OmT (k)/k = E_{\rm T} (k)$
analogously to $\OmM$; see the Appendix.
On the other hand, the first term is
$\partial_{t_{\rm phys}}^2 \tilde{h}_{ij}^{\rm phys} (\kk, t)$
in Fourier space, and $\omega^2
\tilde{h}_{ij}^{\rm phys} (\kk, \omega) = c^2 k^2_{\rm phys}
\tilde{h}_{ij}^{\rm phys} (\kk, \omega)$
if we Fourier transform this term also in time.
This leads again to a similar relation, although now in frequency
space: $k^4 \hc^2 (k, \omega) \sim \int \OmT
(k, t)\,e^{-i \omega t}/t^2\, d t$.
The $\OmGW(k,t)$ spectrum is computed by shell integration
of $\partial_{t_{\rm phys}} \tilde{h}_{ij}^{\rm phys} (\kk, t)\,
\partial_{t_{\rm phys}} \tilde{h}_{ij}^{\rm phys} (\kk, t)$, which
is $c^2 k_{\rm phys}^2 \tilde{h}_{ij}^{\rm phys} (\kk, \omega)
\tilde{h}_{ij}^{\rm phys} (\kk, \omega)$ in the frequency domain.
Hence, we have the asymptotic relation
$\OmGW(k, \omega) \sim k^2 \hc^2(k, \omega)$.
This leads to the observed behavior $\OmGW(k) \sim
\OmT (k)/k^2$ for any fixed time or frequency.

\FFig{pspecm_M1152e_exp6k4} shows a GW spectrum $\OmGW(k)$
that asymptotically falls off faster by a $k^2$ factor than
the magnetic spectrum $\OmM(k)$ in the inertial range.
This is explained by noting that $\OmM(k)$ and $\OmT(k)$
follow the same power law in the inertial range.
Hence, $\OmGW(k) \sim \OmT(k)/k^2 \sim \OmM(k)/k^2$.
For wave numbers below the spectral peak
$k_{\rm GW} \approx 2 k_\ast$,
the GW spectrum, $\OmGW(k)/k$, becomes essentially flat.
This small-$k$ behavior in $\OmGW (k)$
can be traced back to the white noise ($k^2$) spectrum of the magnetic stress
$\ET (k)$,
which seems to emerge even when the magnetic field itself has a
spectrum $\EM (k) = \OmM(k)/k$
steeper than $k^2$ in the subinertial range; see the Appendix.
This argument shows that the scaling of $\OmGW(k)$
with $k^3$ obtained in previous analytical estimates as in,
e.g., Ref.~\cite{Gogoberidze:2007an}, is not expected for the
turbulent developed spectrum.
The subinertial power law $\OmGW (k)\sim k$ is a novel result from
our simulations that was not obtained in previous analytical
estimates.

The characteristic
amplitude has the following asymptotic behavior:
$\hc (k) \sim \OmT^{1/2}(k)/k^2 \sim
\OmGW^{1/2} (k)/k$ for a fixed instant of time.
Looking at Fig.~1 of Ref.~\cite{Gogoberidze:2007an}, we see that
their subinertial range slope in $\hc(k)$ is 1/2.
This slope in $\hc(k)$ corresponds to the +3 slope in $\OmGW(k)$
mentioned above.
At high frequencies, our spectrum $\hc (k)$ has a slope of $-7/3$ corresponding to
a magnetic spectrum of Kolmogorov type.
This agrees with what has been obtained in recent analytic work
\cite{NSS18,SSS19}, although earlier work \cite{Gogoberidze:2007an}
reported a slope of
$-10/3$, which we would obtain if we used a small magnetic Reynolds
number, which results in a $k^{-8/3}$ Golitsyn-type spectrum for the
magnetic field $\OmM (k)$; see \Tab{Tslopes}.

\begin{table}[b!]\caption{
Correspondence between the slopes expected
from Ref.~\cite{Gogoberidze:2007an}
for the subinertial range (``ana'') and what is obtained
in our run~ini2 (``sim''), and
the results for spectra with the Kolmogorov slope
(``Kol'') and the Golitsyn slope (``Gol''),
which agrees with Ref.~\cite{Gogoberidze:2007an}.
}\vspace{12pt}\centerline{\begin{tabular}{cccccccc}
slope of\;\; & ana & sim & Kol & Gol \\
\hline
$\OmM$  &  5  &  5  &  $-2/3$ & $-8/3$  \\
$\OmGW$ &  3  &  1  &  $-8/3$ & $-14/3$ \\
$\hc$  & 1/2 & $-1/2$ &  $-7/3$ & $-10/3$ \\
\label{Tslopes}\end{tabular}}\end{table}

\begin{figure}[t!]\begin{center}
\includegraphics[width=\columnwidth]{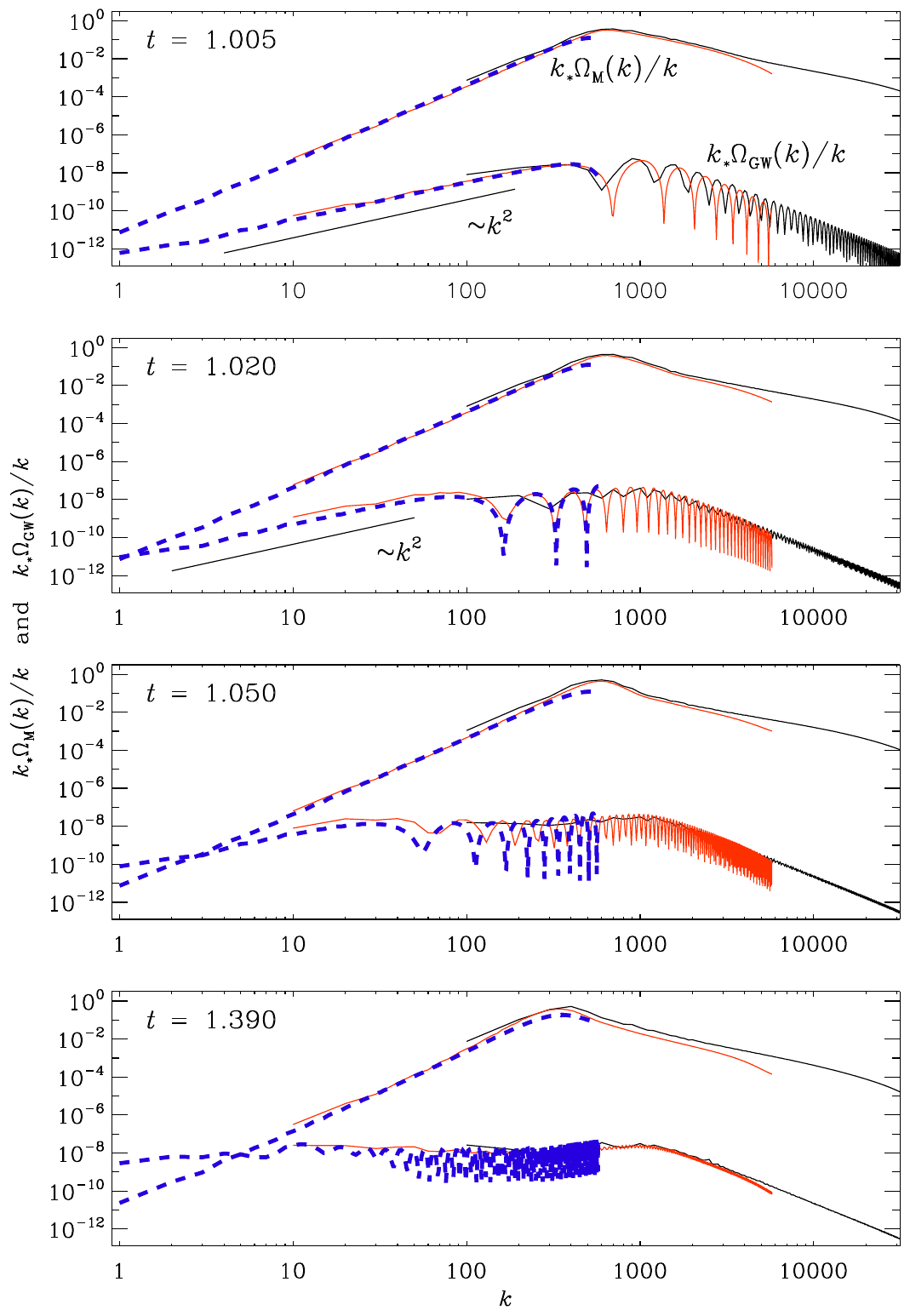}
\end{center}\caption[]{
Time evolution of the magnetic and GW energy spectra
amplified by a factor of $\kf$
for run ini2.
Also shown are the results for a domain larger by a factor of
10 (red) and 50 (blue).
All runs have $1152^3$ mesh points.
}\label{ppower_comp_M1152e_exp6k4}\end{figure}

The emergence of a flat GW spectrum in \Fig{pspecm_M1152e_exp6k4} makes
one wonder how this can be reconciled with the principle of causality.
We recall that the reason for the magnetic energy spectrum to have a
$k^4$ subinertial range is indeed causality combined with the fact that
the magnetic field is solenoidal.
A flat GW energy spectrum would imply that there must have been
GW energy immediately at the largest possible length scales, which would
be unphysical.
Therefore, we have performed numerical simulations with very small time
steps to study how the novel low-$k$ spectrum develops at initial times.
In \Fig{ppower_comp_M1152e_exp6k4}, we show that initially, the
GW spectrum is indeed proportional to $k^2$ and that similar
spectra are also being reproduced in 10 and 50 times larger domains.
It is only during the first few time steps of the numerical simulation
that the GW spectrum is
still proportional to $k^2$,
until the $k^0$ scaling extends over the range between the stirring scale
and the lower wave number in our simulations.
We observe the development of this flat spectrum
for the different sizes of the numerical domain.
This rules out the possibility that this scaling is due to numerical artifacts,
indicating that the flat spectrum is physical and emerges only later.

\begin{figure}[t!]\begin{center}
\includegraphics[width=.95\columnwidth]{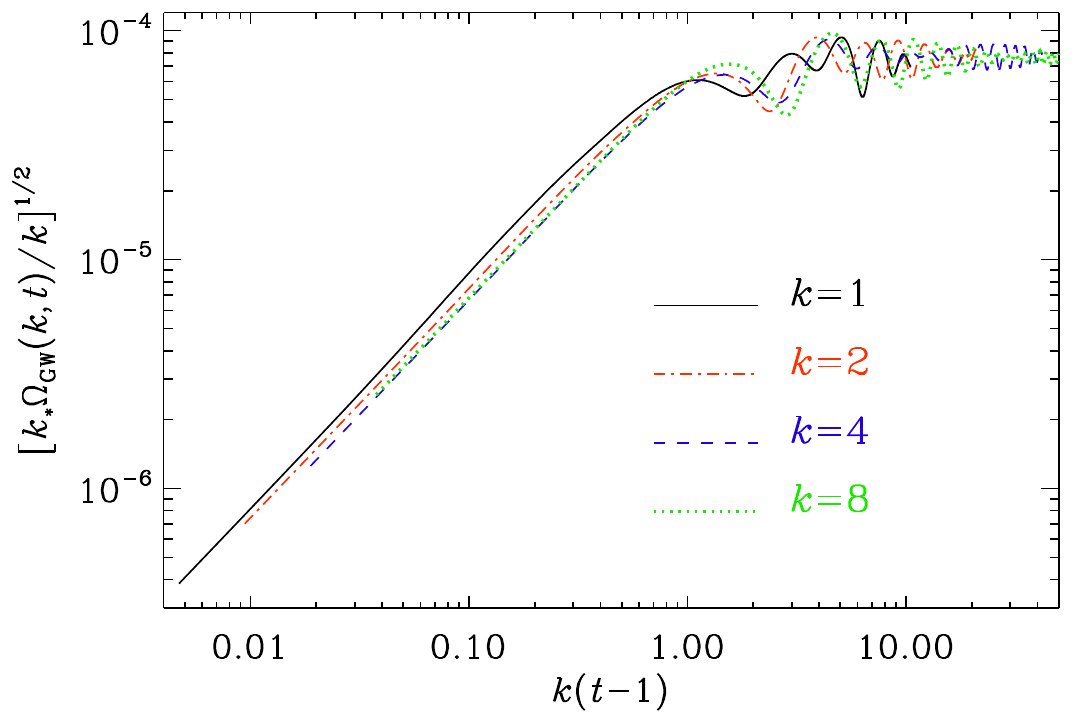}
\end{center}\caption[]{
GW spectral energy versus time for four values of $k$,
demonstrating the $k^2$ scaling at early times for run ini2.
}\label{ppower_comp_M1152e_exp6k4_vs_t}\end{figure}

The time it takes for the change of slope to occur
below the horizon scale is much
shorter than the time it takes for the GW spectrum to
become stationary.
Therefore, we conclude that the $+2$ slope in $\OmGW(k)/k$
is not relevant for the
characterization of the signal.
To demonstrate this further, we show in
\Fig{ppower_comp_M1152e_exp6k4_vs_t} that at small $k$,
$\Omega_{\rm GW}(k,t)/k$ grows with $t$ proportional to $k^2(t-t_*)^2$,
where $t_* = 1$ in normalized units, and reaches
a constant level that is independent of $k$ and is given by the
white noise spectrum of the source at large scales.
This is demonstrated for wave numbers as small as a few times the
Hubble horizon wave number, $k=1$, $2$, $4$, and $8$.

\subsection{Spectra from the electroweak phase transition}

\begin{figure}[t!]\begin{center}
\includegraphics[width=\columnwidth]{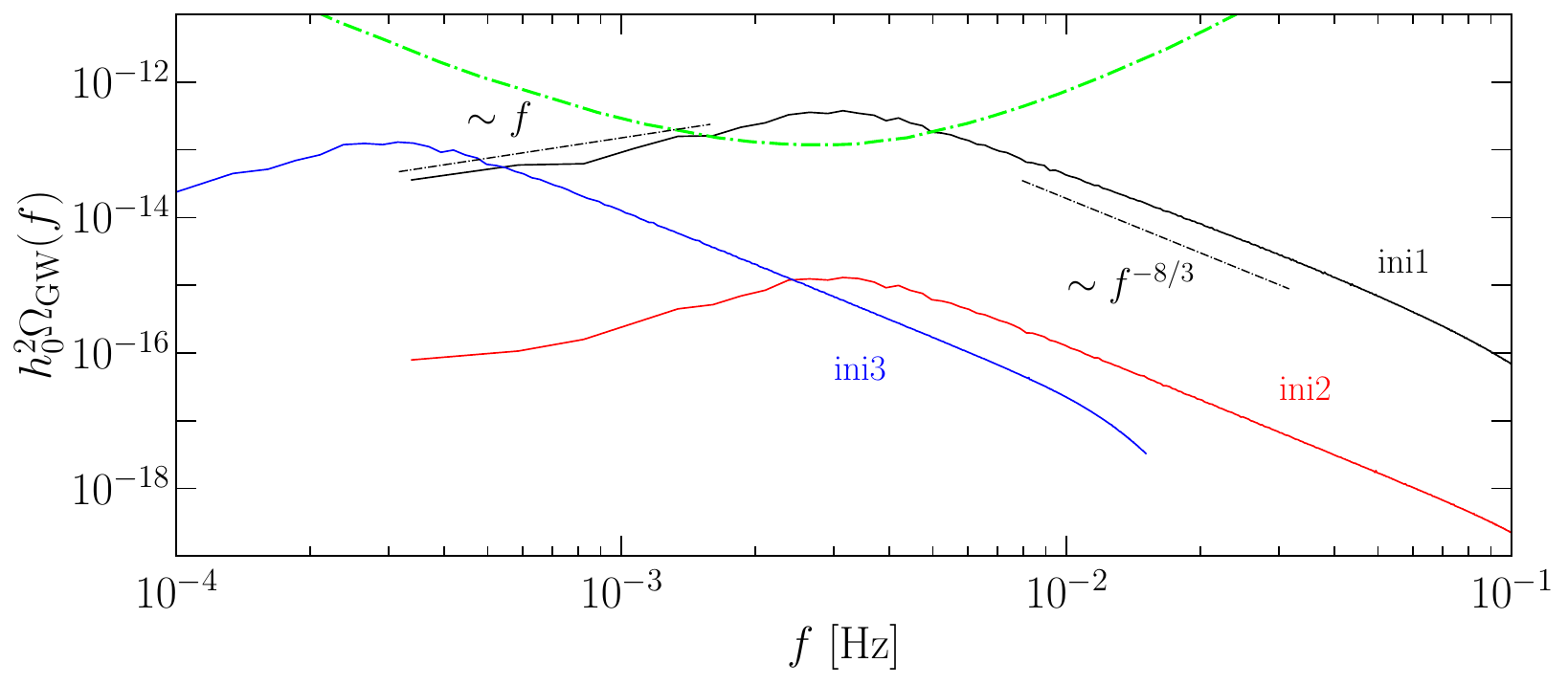}
\includegraphics[width=\columnwidth]{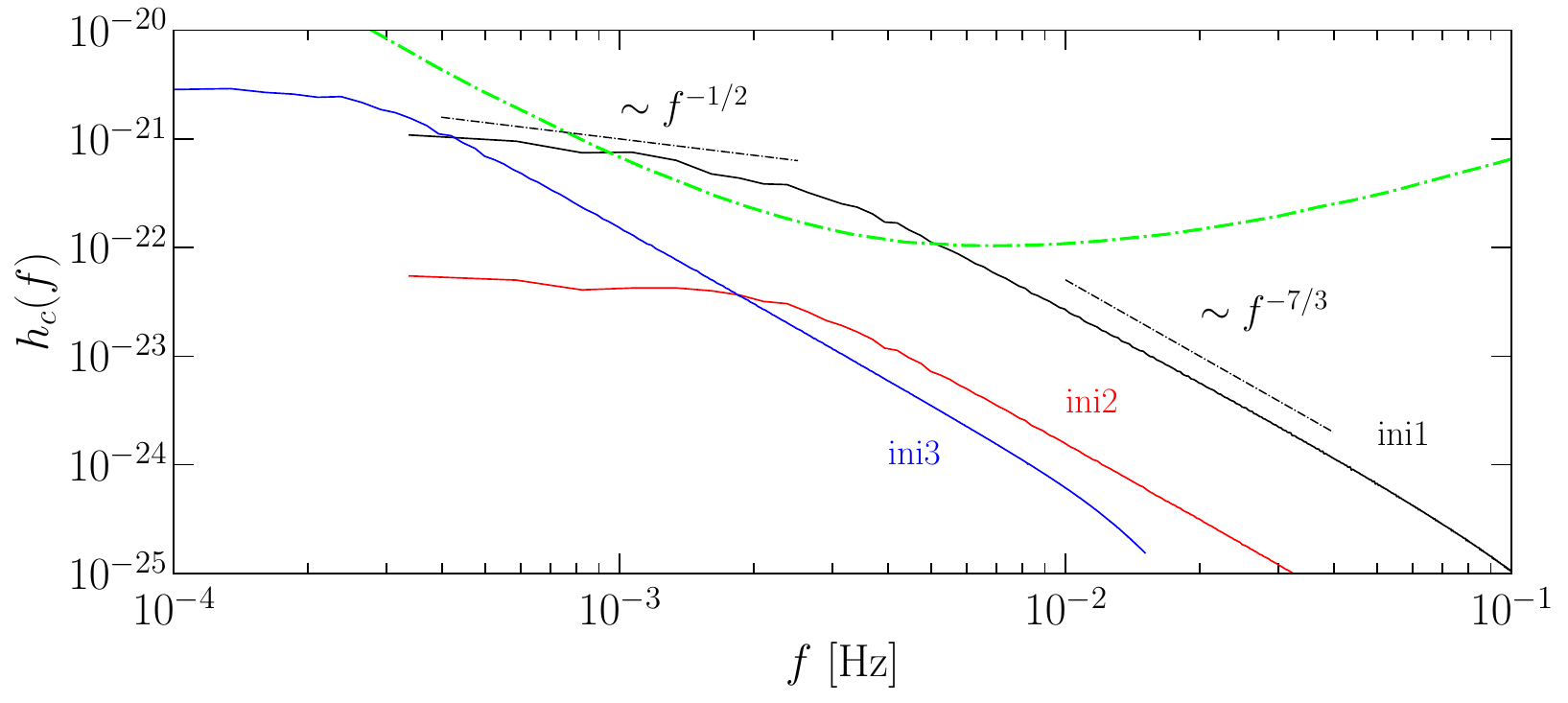}
\end{center}\caption[]{
Spectra of $h_0^2\OmGW(f)$ and $\hc(f)$ evaluated at the present time,
along with the LISA
\blue{power law} sensitivity curve (green dot-dashed line) to a stochastic GW background
\blue{assuming} four years of mission, \blue{and a threshold signal-to-noise ratio of 10
 \cite{RCC18, SC19,Caprini:2019pxz}}. See \Tab{Tsummary} for details of runs ini1--3.
}\label{pspecm_scl5_comp2}\end{figure}

The GW energy density
$h_0^2 \OmGW (f)$ and the characteristic
strain amplitude $\hc (f)$
are shown in \Fig{pspecm_scl5_comp2} for runs~ini1--3
as a function of the frequency $f$,
all shifted to the present time
as defined in \Sec{GW_eq}.
These are obtained
by scaling the computed normalized GW spectra to the
physical spectra produced at the electroweak
scale.
If we take $T_\ast = 100 \GeV$ for the temperature,
and $g_\ast \approx g_{\rm S} = 100$ for the number of
relativistic and adiabatic degrees of freedom at the electroweak
phase transition, the Hubble rate is [see \Eq{rho_rad}]
\begin{equation}
H_* \approx 2.066 \times 10^{10} \s^{-1} \left(k_{\rm B}
T_* \over 100 \GeV \right)^2
\left( g_* (T_*) \over 100 \right)^{1/2},
\label{H*}
\end{equation}
where the proportionality factors $T$ and $g_\ast$ are kept as a parameter
due to the uncertainty of the exact values.
Our convention of setting $a_\ast = 1$ leads to the following value of
$a_0$ computed assuming adiabatic expansion of the Universe [see
text above \Eq{EEGW}],
\begin{equation}
a_0 \approx 1.254 \times 10^{15} \left(\kB T_* \over 100 \GeV \right)
\left( g_{\rm S} (T_*) \over 100 \right)^{1/3},
\label{a0}
\end{equation}
where we have used the values $g_{\rm S} = 3.91$ and $T_0 = 2.73\K$
at the present time.

For different scenarios, the results scale in the following way:
The frequency shifts proportional to $T g_{\ast}^{1/2} g_{\rm S}^{-1/3}$,
the strain amplitude varies with $T^{-1} g_{\rm S}^{-1/3}$,
and the GW energy density with $g_\ast g_{\rm S}^{-4/3}$.
Likewise, the
GW strain amplitude is proportional to the stirring scale
$N^{-3/2}$ and the frequency is
proportional to $N$ \cite{Gogoberidze:2007an}.

The slopes of $\OmGW (f)$ in \Fig{pspecm_scl5_comp2}
are consistent with those
in \Fig{pspecm_M1152e_exp6k4}, where $\OmGW (k)$ was shown,
due to the dispersion relation; see \Eq{freq}.
The spectrum of $\hc(f)$ shows scaling with $f^{-1/2}$ for low frequencies and
with $f^{-7/3}$ in the inertial range,
as expected.
As discussed above, we expect the subinertial
slope of $-1/2$ to eventually turn over a slope
of $+1/2$ as $f$ and time decrease, due to the lack of causality on scales
larger than the horizon.
However, the simulation domains are smaller
than the horizon scale, so this turnover is not observed.

The analytic approximation in Ref.~\cite{Gogoberidze:2007an}
gives a peak value $\hc \approx 4\times10^{-20}$ at $1 \mHz$
for their largest Mach number of unity (see Fig.~1 of
Ref.~\cite{Gogoberidze:2007an}).
By comparison, for our run~ini1, the spectrum shows an intermediate peak
at $3\mHz$ with \blue{$\hc\approx0.7\times10^{-21}$}; see \Fig{pspecm_scl5_comp2}.

\begin{figure}[t!]\begin{center}
\includegraphics[width=\columnwidth]{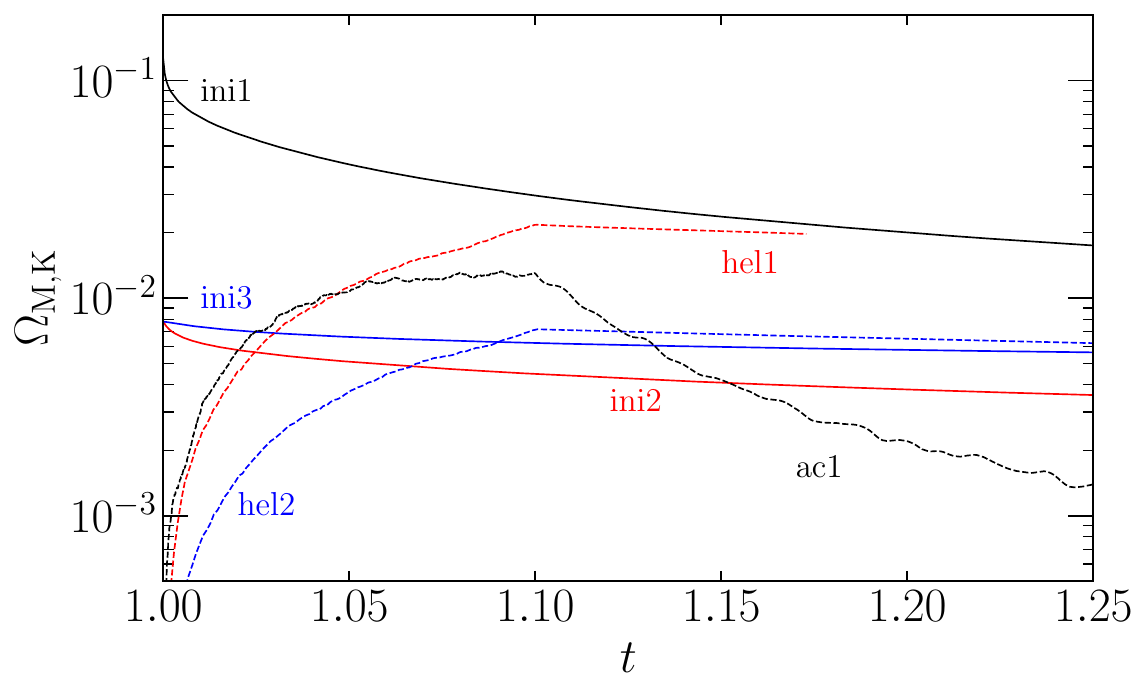}
\includegraphics[width=\columnwidth]{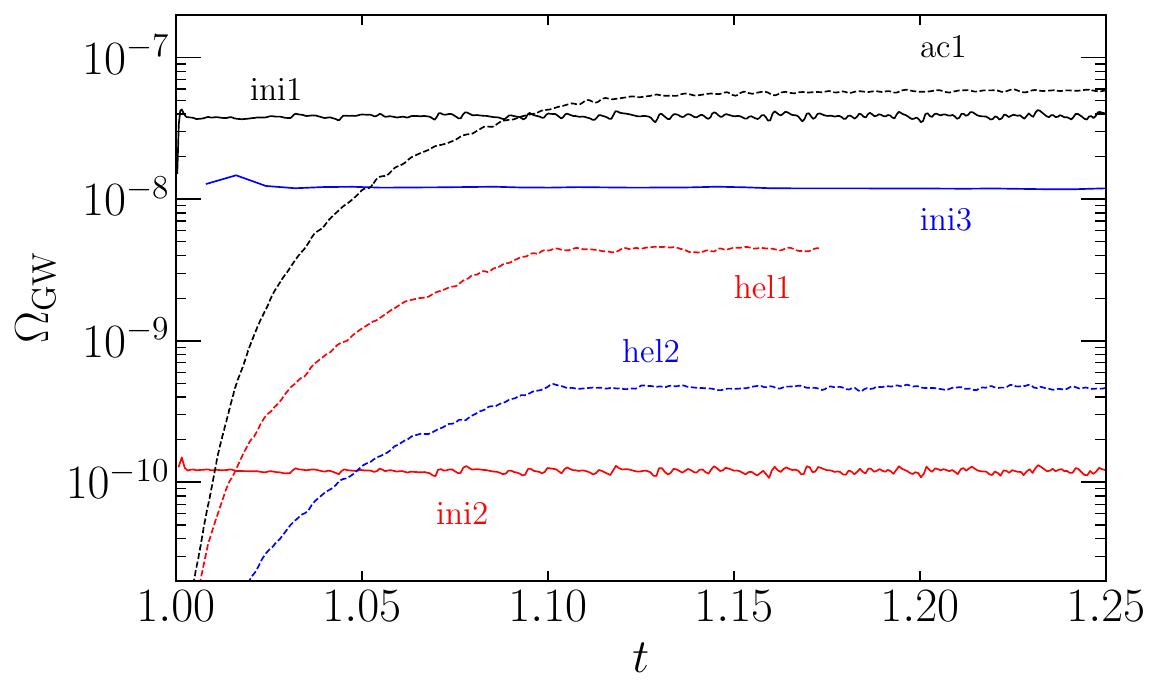}
\end{center}\caption[]{
Evolution of $\Omega_{\rm M, K}$ (top) and $\OmGW$ (bottom) for
runs with initial energy (ini1--3) and runs where energy is driven
through monochromatic forcing (hel1--2 and ac1).
Note that the energy densities are normalized with the radiation
energy density at the time of generation.
}\label{pcomp_EEGW}\end{figure}

In runs~ini1--3, GWs are produced by the sudden emergence of a magnetic field.
In reality, this will be a gradual process, as modeled by sets~II and III of runs.
The time evolution of $\Omega_i$ (for $i=\,$GW, K, or M)
integrated over all wave numbers
is shown in \Fig{pcomp_EEGW} for ini1--3, hel1--2, and ac1.
In all these cases, the GW energy density saturates at a value
$\OmGW^{\rm sat}$ shortly after the sourcing energy density
has reached its maximum value $\Omega_{\rm M, K}^{\rm max}$; see
\Tab{Tsummary}.

\begin{figure}[t!]\begin{center}
\includegraphics[width=\columnwidth]{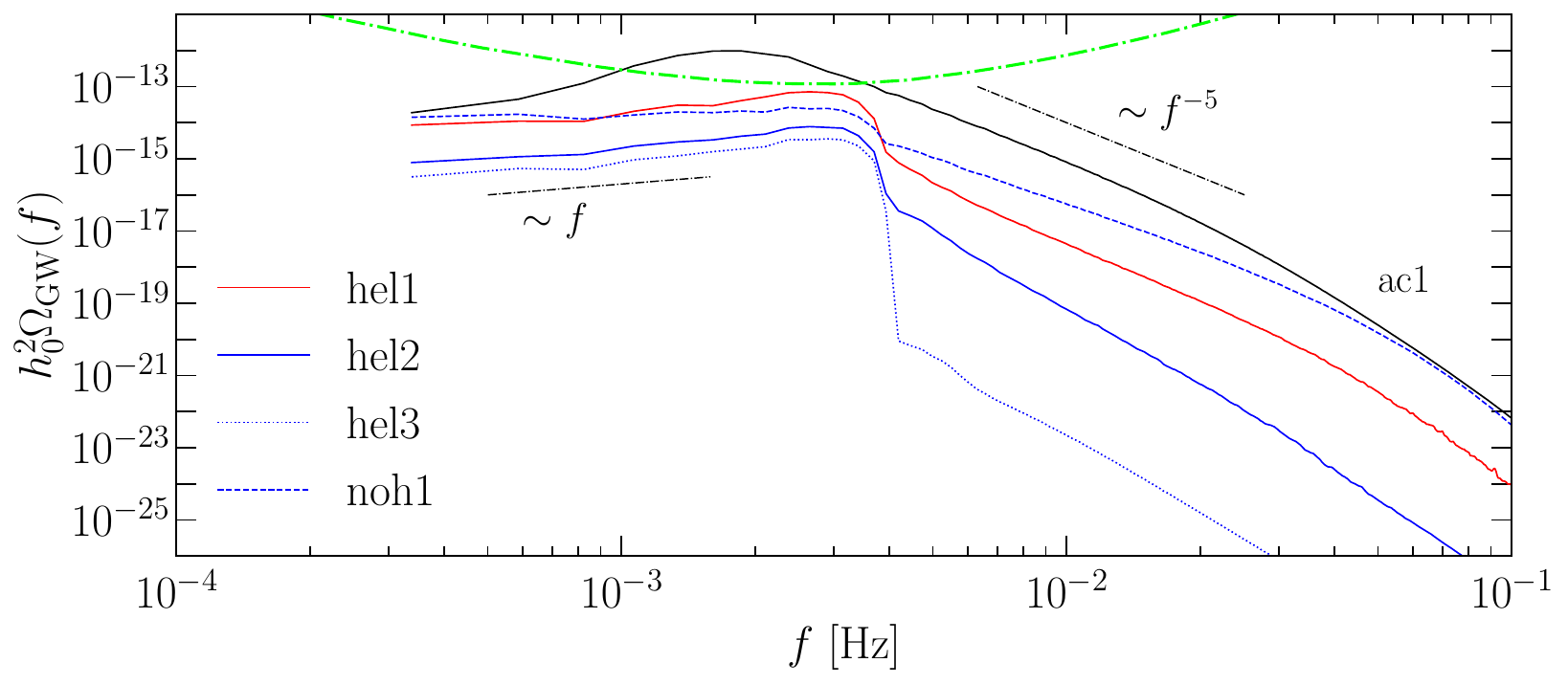}
\includegraphics[width=\columnwidth]{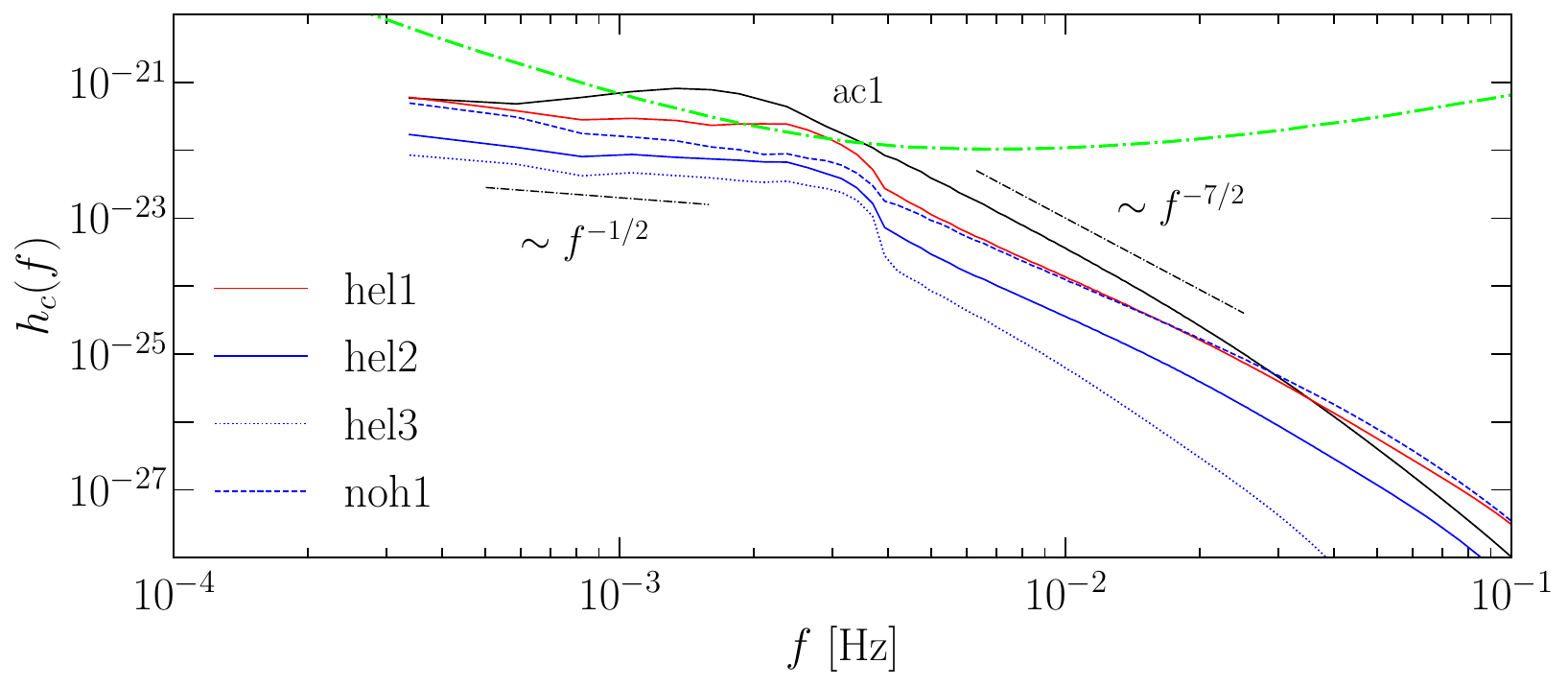}
\end{center}\caption[]{
Similar to \Fig{pspecm_scl5_comp2}, but for runs~hel1--3, noh1, and ac1.
}\label{pspecm_scl5_comp}\end{figure}

In \Fig{pspecm_scl5_comp}, we observe that runs~hel1--3, noh1, and ac1 present
steeper GW spectra at high frequencies than in runs~ini1--3.
The monochromatic forcing produces a spike in $\OmM (f)$ at $f_\ast$
and a sharp drop in $\OmGW (f)$ and $\hc(f)$ beyond $f_{\rm GW}
\approx 2 f_\ast$, for the magnetic runs.
For the acoustic runs, we observe a smooth bump on the spectra
of $\OmGW(f)$ and $\hc(f)$ near the spectral peak $f_\ast$.
Again, the spectra have the same low-frequency tail,
which underlines its universal nature.
Also, for a given input energy, $\Omega_{\rm M, K}^{\rm max}$, we
obtain larger values of $\OmGW^{\rm sat}$ for acoustic than
for vortical turbulence.
This case was already studied in Refs.~\cite{HHRW15,HHRW17,NSS18}.
These features could not be captured by previous analytical estimates,
and the power laws in the inertial range also seem to be affected by
how the turbulent fields are driven at initial times.

\begin{figure}[t!]\begin{center}
\includegraphics[width=\columnwidth]{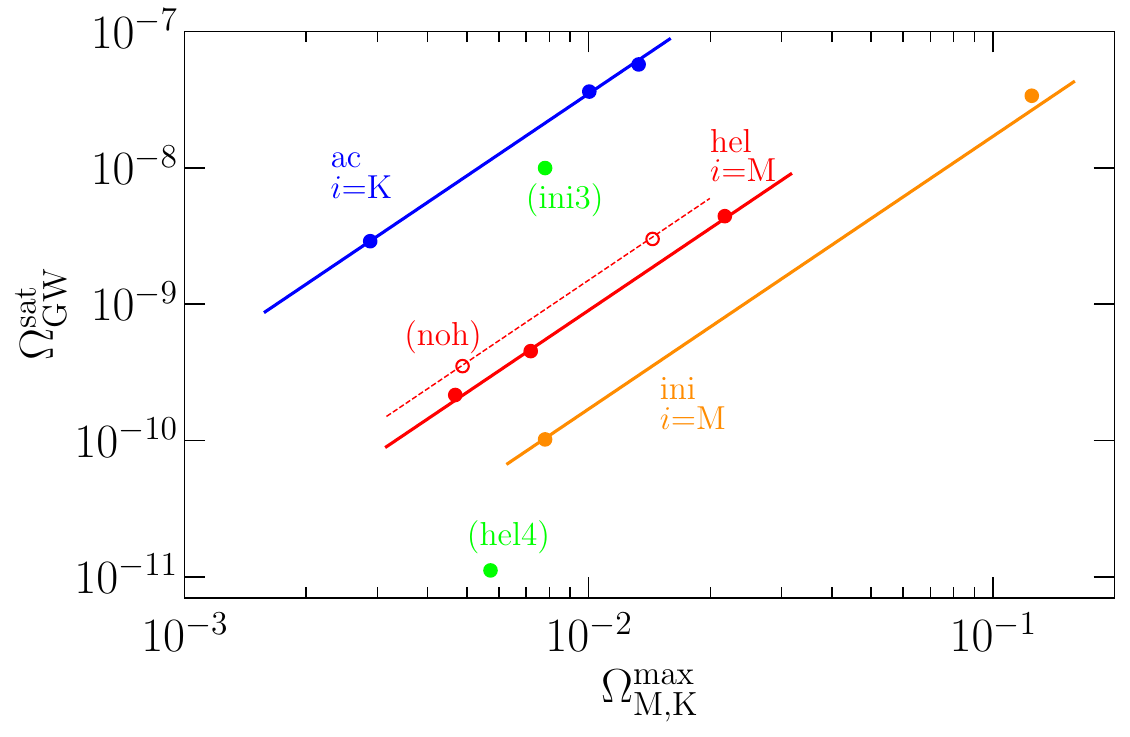}
\end{center}\caption[]{
$\OmGW^{\rm sat}$ versus
$\Omega_{\rm M, K}^{\max}$.
The quadratic dependence inferred from the $+2$ slope of the lines
holds within runs of the same type.
Note that runs~ini3 ($N = 10$) and hel4 ($N = 1000$) in green
have different stirring scales than the rest of the runs ($N = 100$).
}\label{EEGW_vs_EEKM}\end{figure}

For a given type of initial condition and stirring scale, the final energy density in
GWs has the expected quadratic dependence on the source
energy density to a very good approximation
as shown in \Fig{EEGW_vs_EEKM}.
The efficiency of
GW production varies significantly with the type of initial conditions;
for the same total source energy,
the cases with forced acoustic compression
lead to  a factor of around \blue{$200$} more GW energy than those with a sudden
magnetic field (ini1--3), while the cases with forced
nonhelical magnetic fields are \blue{$10$} times more efficient than the latter.
We also observe that nonhelical forcing fields are about a
factor of $1.6$ more efficient than helical magnetic fields.
The detailed reasons behind these significant variations in efficiency
are unclear, but they imply that accurate predictions of GW production from
cosmological phase transitions will require a detailed model of how latent
heat is converted to plasma and magnetic field energies.
The comparison of efficiency in GW generation between acoustic and
rotational turbulence is a
subject of further investigation.

\subsection{LISA Detectability}
\label{prospects}

The projected sensitivity curve for the LISA space mission
was plotted in \Figs{pspecm_scl5_comp2}{pspecm_scl5_comp} along with
GW spectra from our runs.
The plotted sensitivity assumes a mission of four years \cite{RCC18, SC19},
\blue{and corresponds to the power law sensitivity obtained assuming a threshold
signal-to-noise ratio (SNR) of 10 \cite{Caprini:2019pxz}}.
The cases ini2--3, each with a turbulent energy input of around 1\% of the total radiation
energy density, produce \blue{signals with a SNR below 10},
while ini1, with a turbulent energy input of around 10\%,
could be detectable.
An energy input of about \blue{8\%} is required to obtain a GW spectrum \blue{with a
SNR above 10} for runs with an initial helical magnetic field \blue{when $N=100$,
and 3\% when $N=10$.}
\blue{On the other hand}, the runs with forced magnetic fields would \blue{be detectable}
for an energy input of approximately \blue{4\%},
according to our results.
Acoustic forced turbulence has been shown to be the more efficient case considered,
even though it leads to a GW spectral peak closer to the forcing peak, which
slightly reduces the prospects of detection for $T_\ast = 100 \GeV$ and $N = 100$.
An energy input of around \blue{0.5\%} would be enough in this case for \blue{the GW
signal to have a SNR above 10}.
\section{Conclusions}
\label{conclusions}

In the early universe, hydrodynamic and MHD turbulence can be an efficient
source of GWs.
Our direct numerical simulations have shown that the GW energy produced by
the turbulence depends quadratically on the energy of the turbulence at
the time turbulence is strongest.
In the inertial range of the turbulence, the slope of the GW spectrum is
steeper than the slope of the magnetic energy spectrum by a factor of $k^2$.
For a magnetic energy spectrum of Kolmogorov type of the form $\OmM\sim
k^{-2/3}$, the GW energy spectrum is of the form $\OmGW\sim k^{-8/3}$.
In the subinertial range, however, where the magnetic energy spectrum
is expected to be proportional to $k^5$, the GW energy spectrum is not
proportional to $k^3$, as naively expected, but proportional to $k$.

The shallow subinertial range spectrum for the GW energy also implies a
novel $f^{-1/2}$ low-frequency spectrum for $\hc(f)$.
This would enhance the detectability of such a signal
compared to the $f^{1/2}$ spectrum obtained from
earlier analytic models typically assumed in recent analyses as, e.g., in Ref.~\cite{croon18}.

\begin{figure*}[t!]\begin{center}
\includegraphics[width=0.9\textwidth]{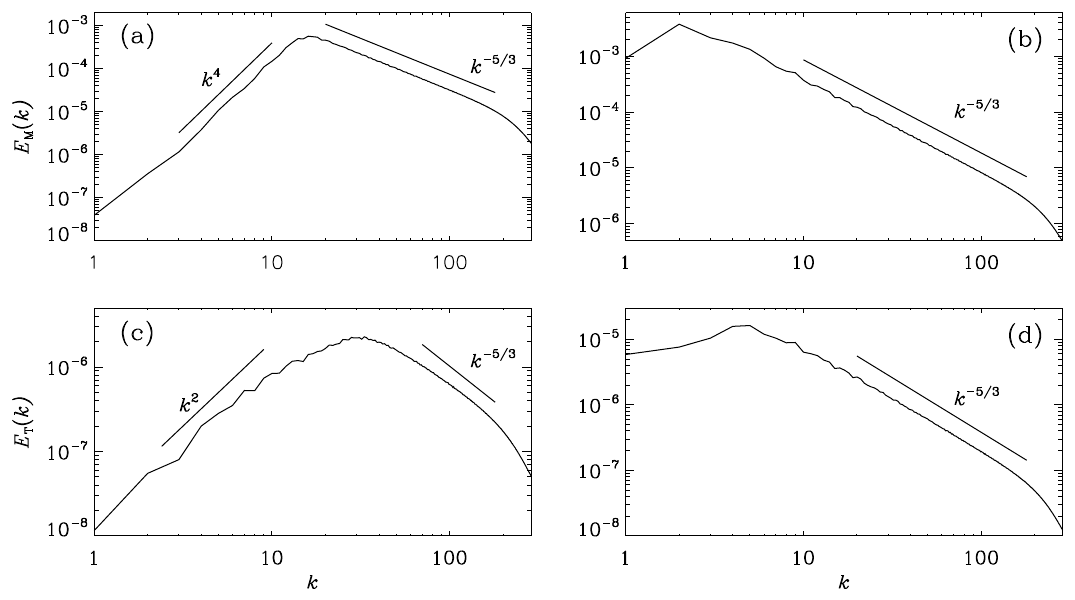}
\end{center}\caption[]{
Shell-integrated spectra of the vector $\BB$, $\EM (k)$, and of the scalar $\BB^2$,
$\ET (k)$, for a random nonhelical magnetic field, with spectral peak
at $\kf = 15$ (left panels), and $\kf = 2$ (right panels).
We see that the magnetic spectrum $\EM (k)$ has the same slope in the inertial range
as that of the stress spectrum $\ET (k)$.
In the subinertial range, when $\BB$ has a Batchelor $k^4$ spectrum,
the spectrum of $\BB^2$ is always that of white noise, i.e., proportional to $k^2$.
}\label{pspec_sqr_comp}\end{figure*}

Comparing vortical MHD turbulence with irrotational turbulence driven by
spherical expansion waves, we find that at similar turbulent energies,
irrotational turbulence appears to drive GW energy more efficiently than
vortical MHD turbulence.
This may be connected with the temporal correlations of the turbulence.
Depending on the specific dynamical evolution during the symmetry breaking process,
the GWs produced by primordial turbulence may
be detectable with LISA when the fraction of radiation energy converted
into turbulent energy exceeds a value between \blue{0.5\%} and 10\%.
In addition to the total GW energy density,
the spectral shape is also affected by the dynamical evolution
of the magnetic and/or velocity fields during the phase transition.
The specific features around the spectral peak
and the power law in the inertial range require
numerical simulations to be accurately described.

Now scheduled for launch in the mid 2030s,
LISA may provide crucial insight into fundamental physics during
the first picoseconds of cosmic evolution.

\vspace{2mm}
Data availability---The source code used for the
simulations of this study, the {\sc Pencil Code},
is freely available from Ref.~\cite{PC}.
The simulation setup and the corresponding data are freely available from
Ref.~\cite{DATA}.

\acknowledgements
\vspace{3mm}

Support through the NSF Astrophysics and Astronomy Grant Program
(Grants No.~1615940 and No.~1615100) and the Shota Rustaveli NSF
(Georgia) (Grant No.~FR/18-1462) are gratefully acknowledged.
We acknowledge the allocation of computing resources provided by the
Swedish National Allocations Committee at the Center for Parallel
Computers at the Royal Institute of Technology in Stockholm.

\vspace{4mm}
\section*{Appendix: Spectrum of the source}
\label{B2spec}
\vspace{1mm}

We compare in \Fig{pspec_sqr_comp} the shell-integrated
spectrum of the magnetic field,
$\EM (k) = \OmM(k)/k$ defined such that $\int_0^{\infty}
\EM (k) \, d k = \int_{-\infty}^\infty \OmM (k) \, d\ln k$,
with the spectrum of the stress tensor $T_{ij}$.
The latter spectrum
$\ET (k) =  \OmT (k)/k$ is defined
such that $\int_0^\infty \ET (k) \, dk = \bra{T_{ij} T_{ij}}/2$.
Note that, in the absence of fluid motions, this corresponds
to the spectrum of the squared magnetic field, 
whose integral over all wave numbers gives $\bra{(\BB^2)^2}/2$
instead of just $\bra{\BB^2}/2$.

In the $k$ range where $\EM (k)$ has a Batchelor ($k^4$) spectrum,
the stress spectrum $\ET (k)$ is white noise ($k^2$).
The $k^4$ spectrum is caused by a white noise ($k^2$) spectrum of the vector potential.
Recent work \cite{BB19} shows that the spectrum of the stress
$\ET (k)$ can never be steeper than that of white noise, and
that the peak of the stress spectrum shifts to $2\kf$,
being $\kf$ the position of the spectral peak of the magnetic field,
as it is observed in \Fig{pspec_sqr_comp}.
In the inertial range, we observe both spectra $\EM (k)$ and $\ET (k)$
to possess the same Kolmogorov scaling $k^{-5/3}$.
These results have also been confirmed analytically by calculating
the spectrum of the stress as the autocorrelation function of the two
turbulence spectra under the assumption that the underlying fields are
Gaussian distributed \cite{BB19}.
As inferred from \Eq{GW4}, the relevant spectrum related to the
GW energy is the stress spectrum, instead of the magnetic spectrum.


\end{document}